\documentclass[11pt,draftclsnofoot,onecolumn,journal]{IEEEtran}
\usepackage{amsthm,amsmath,amssymb}
\IEEEoverridecommandlockouts
\usepackage[dvips]{graphicx}
\usepackage[dvips]{color}  
\usepackage{mathrsfs}
\usepackage{subfigure}

\DeclareMathAlphabet{\mathbit}{OT1}{cmr}{bx}{it}
\def\Rset{\mathrm{ I\mkern-3.5mu R}}

\newtheorem{lemma}{Lemma}
\newtheorem{theorem}{Theorem}
\newcommand{\st}{\rule[-1mm]{0mm}{4mm}}

\title{Coding Theorems for Repeat Multiple Accumulate Codes}
\author{\authorblockN{J{\"o}rg Kliewer\authorrefmark{1}, Kamil
    S.~Zigangirov\authorrefmark{3}, Christian Koller\authorrefmark{3},
    Daniel J.~Costello, Jr.\authorrefmark{3}\\[2ex]}
  \authorblockA{\authorrefmark{1}Klipsch School of
    Electrical and Computer Engineering,\\
    New Mexico State University, Las Cruces, NM 88003, USA\\
    Email: jkliewer@nmsu.edu\\[2ex]}
  \authorblockA{\authorrefmark{3}Department of Electrical Engineering,\\
    University of Notre Dame, Notre Dame, IN 46556, USA\\
    Email: \{kzigangi, ckoller, costello.2\}@nd.edu} \thanks{This work
    was supported in part by NSF Grants CCF-0830666 and CCF-0830650,
    NASA Grant NNX07AK53G, and the University of Notre Dame Faculty
    Research Program. Parts of the paper were presented at the 45th
    Annual Allerton Conference on Communication, Control, and
    Computing, September 2007, Monticello, IL \cite{KZC07}.}  \vspace*{-7ex}}

\begin{document}

\maketitle

\begin{abstract}
\vspace*{-1.5ex}
  In this paper the ensemble of codes formed by a serial
  concatenation of a repetition code with multiple accumulators connected
  through random interleavers is considered. Based on finite length weight
  enumerators for these codes, asymptotic expressions for the minimum
  distance and an arbitrary number of accumulators larger than one are
  derived using the uniform interleaver approach. In accordance with 
  earlier results in the literature, it is first shown that the minimum distance
  of repeat-accumulate codes can grow, at best, sublinearly with block length.
  Then, for repeat-accumulate-accumulate codes and rates of $1/3$ or less, 
  it is proved that these codes exhibit asymptotically linear distance 
  growth with block length, where the gap to the Gilbert-Varshamov bound can
  be made vanishingly small by increasing the number of accumulators beyond two. 
  In order to address larger rates, random puncturing of a low-rate mother
  code is introduced. It is shown that in this case the resulting ensemble of 
  repeat-accumulate-accumulate codes asymptotically achieves linear distance 
  growth close to the Gilbert-Varshamov bound. This holds even for very 
  high rate codes.

\end{abstract}

\begin{keywords}\vspace*{-1.5ex}
Multiple serial concatenation, repeat-accumulate codes, uniform
interleaver, minimum distance growth rate coefficient, Gilbert-Varshamov bound
\end{keywords}

\section{Introduction}
Since the invention of turbo codes, several new turbo-like coding
schemes have been proposed. Among these
are serially concatenated codes, a simple example of which is the
repeat-accumulate (RA) code \cite{DJM98} consisting only of a
repetition code, an interleaver, and an accumulator.  Such a code has
the advantage of very low decoding complexity compared to serially
concatenated code constructions with more complex constituent codes.
Another benefit of RA codes, compared to powerful code constructions
such as LDPC codes, is their extremely low encoding complexity of
$O(1)$, whereas LDPC codes have an encoding complexity of $O(g)$,
where $g$, although much smaller than the block length \cite{RU01}, is
greater than one.  This makes RA codes well suited in power-limited
environments; for example, for physical layer error correction in
battery powered sensor network nodes or for spacecraft communications.

Design guidelines for achieving an interleaver gain with double
serially concatenated code constructions have been given in
\cite{BDMP98a} (and in \cite{JM02} for more general parallel and
serially concatenated code ensembles).  In this paper we address
multiple serially concatenated RA-type codes, where, in particular, 
we focus on the serial concatenation of an outer repetition code with 
two or more accumulators connected through random interleavers. The 
resulting code ensemble is then analyzed using the uniform interleaver 
approach \cite{BDMP98} by averaging over all possible interleavers. 
Our work is mainly motivated by \cite{PS00,PS03}, where a similar setup 
was considered.  Whereas \cite{PS00} considers the interleaver gain of
repeat multiple accumulate (RMA) codes (and thus addresses a special case 
of the work in \cite{JM02}), in \cite{PS03} the authors show that, for an
asymptotically large number of accumulators, there are codes in the
ensemble whose minimum distance grows linearly with block length and
which achieve the Gilbert-Varshamov bound (GVB).  They also provide
numerical calculations of the minimum distance for different numbers
of accumulators and finite block lengths, but they do not give results
for the asymptotic minimum distance growth rate coefficient in the 
practically more relevant case of a finite (small) number of accumulators.
An extension of this work was presented in \cite{Pfi03},
where it is shown that, for a finite number of accumulators larger
than one, linear minimum distance growth can be obtained, but an
asymptotic growth rate coefficient for the minimum distance was only 
conjectured. This  was also shown in \cite{BMS03} for the special case 
of two accumulators, but again an asymptotic growth rate coefficient 
was not given.

\begin{figure}[htb]
  \centerline{\includegraphics[scale=0.6]{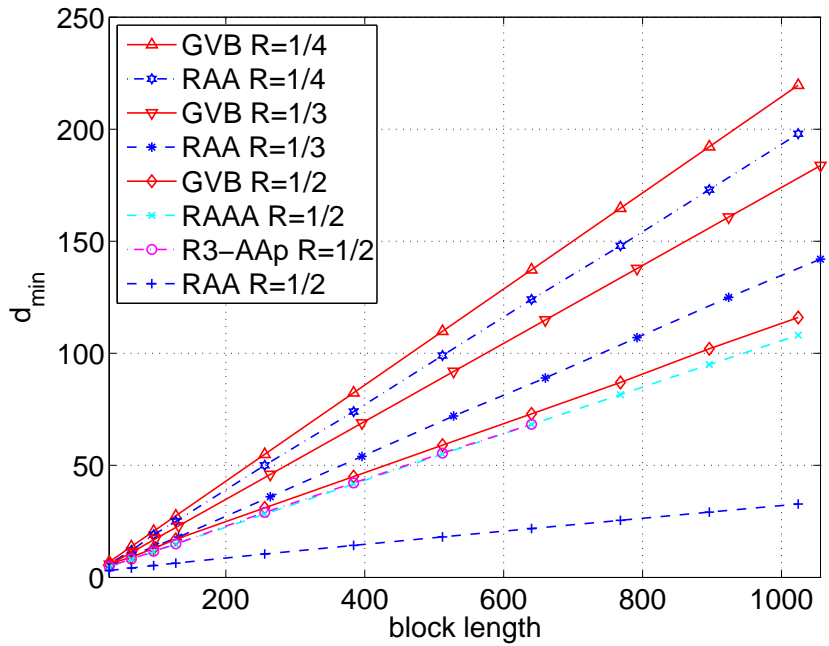}}
  \caption{Lower bounds on the average minimum distance growth rate
    coefficient for finite block lengths for RAA and RAAA code
    ensembles along with the GVB.}
    \label{fig:finite}
\end{figure}

Lower bounds on the average minimum distance growth for \emph{finite}
block lengths are given in Fig.~\ref{fig:finite} for different code
rates $R$ and two (RAA) and three accumulators (RAAA) along with the
GVB. The bounds are computed such that half of the codes in each
ensemble have a minimum distance of at least $d_{\text{min}}$.  We
observe that, for rate $R=1/2$, the growth rate coefficient of the RAA
code is much smaller than the GVB; however, both the RAAA code and the
RAA code punctured from a rate $1/3$ mother code (R3-AAp) have a
minimum distance growth rate coefficient very close to the GVB. Also,
it can be seen that, for code rates of $R=1/3$ and $R=1/4$, the
minimum distance growth rate coefficient is closer to the GVB than in
the $R=1/2$ case. For the rate $R=1/3$ and $R=1/4$ RAAA code ensembles
the minimum distance growth rate coefficients coincide with the GVB
and are therefore not shown in Fig.~\ref{fig:finite}. In contrast to
the results given in Fig.~\ref{fig:finite}, we are interested in the
asymptotic behavior of the minimum distance for RMA codes with block
lengths tending to infinity. In particular, an asymptotically "good"
code ensemble has the property that, for large block lengths, the
minimum distance grows linearly with block length, which cannot be
reliably determined from the finite length analysis shown in
Fig.~\ref{fig:finite}.

In the following, we extend previous results in the literature 
\cite{PS03,Pfi03,BMS03} and present an analysis that fully characterizes
the asymptotic minimum distance behavior of RA and RMA code ensembles.
The main result of the paper is that, for RAA codes
and rates equal to $1/3$ or smaller, these code ensembles exhibit
linear distance growth with block length, where the gap to the GVB can
be made arbitrarily small by increasing the number of accumulators
beyond two.  In addition, we consider random puncturing at the output
of the last accumulator. We show that in this case the resulting
ensemble of RAA codes achieves linear distance growth close to the GVB
for any code rate smaller than one.

The paper is organized as follows. In Section~II we consider the
ensemble average weight spectrum of RA codes and show that the minimum
distance grows as a fractional power of the  block length. Section~III 
addresses the asymptotic minimum distance analysis of a double
serially concatenated RAA code. This concept is extended to
multiple serial concatenation with more than two accumulators in
Section~IV. Finally, random puncturing and its effect on minimum
distance is discussed in Section~V, and some concluding remarks are
given in Section~VI.

\section{Ensemble average weight spectrum for repeat-accumulate codes}
\label{sec:RA}

In this section we briefly address the minimum distance of RA codes
and  show that these codes cannot achieve linear distance growth
with block length, i.e., they are not asymptotically good code
ensembles.  Related results have already been established in
\cite{Pfi03}, \cite{BMS03}, \cite{KU98}, and \cite{PB06}, where lower
and upper bounds on minimum distance for more general serially
concatenated codes have been derived. We restate these results for
tutorial reasons and since we will use them in Section~\ref{sec:RAA},
where we analyze the RAA code ensemble.

\begin{figure}[htb]
  \centerline{\includegraphics[scale=0.8]{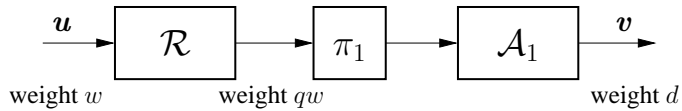}}
    \caption{Repeat-accumulate encoder.}
    \label{fig:RA}
\end{figure}

The RA encoder is shown in  Fig.~\ref{fig:RA}.
The binary input sequence $\mathbit{u}$ has length $K$ and Hamming
weight $w$, and ${\cal R}$ denotes the repetition code of rate
$R=1/q$, which leads to a codeword of weight $qw$ and length $N=qK$.
The subsequent interleaver $\pi_1$ permutes the symbols of the 
codeword.  We consider the ensemble of all interleavers by using the
uniform interleaver approach \cite{BDMP98}, where each possible
interleaver realization has probability $1/N!$. The permuted
output sequence is applied to the recursive convolutional code 
${\cal A}_1$ with generator polynomial $g(D)=1/(1+D)$ (accumulator),
leading to an output sequence $\mathbit{v}$ of weight $d$.

We will characterize RA code ensembles by their input-output weight 
enumerating function (IOWEF) $A_{d,w}$, which is the number of codewords 
having weight $d$ that result from input sequences of weight $w$. 
Let $E(A_{d,w})$ denote the expected value of the IOWEF using the
uniform interleaver approach.

We also define the ensemble-average weight enumerating function (WEF)
$E(A_{d})$, the expected number of codewords of weight $d$, as
\begin{equation}
  E(A_{d}) \triangleq \sum_{w=1}^{K} E(A_{d,w}),
\label{eq:WEF}
\end{equation}
and similarly the ensemble-average cumulative WEF $E(A_{d \le \delta})$ 
specifying the expected number of codewords in the ensemble with 
weight not exceeding $\delta$,
\begin{equation}
  E(A_{d \le \delta}) \triangleq \sum_{d=1}^{\delta} E(A_{d}).
\label{eq:cum_wef}
\end{equation}

For a given code, 
the minimum distance $d_{\text{min}}^{\text{RA}}$ is defined as the smallest
value of $\delta$ for which $A_{d \le \delta}$ is nonzero.

\begin{theorem}
\label{th:RA_lowerbound}
In the ensemble of RA codes with block length
$N\rightarrow \infty$ and $q \geq 3$, almost all codes have minimum distance 
$d_\text{min}^{\text{RA}}$ lower bounded by the inequality 
\[
d^{\text{RA}}_{\text{min}}> N^{\frac{q-2}{q}-\epsilon},
\]
where $\epsilon$ is any fixed positive value. 
\end{theorem}

\begin{proof}
  The conditional probability that a weight $d$
  codeword is obtained at the output of the accumulator for a given 
  input weight $w$ can be expressed as \cite{DJM98,BDMP98}
\begin{equation}
\Pr(d|w)= \begin{cases}
1, & \quad \text{for}\ w=0,\ d= 0,\\
0, & \quad \text{for}\ w=0,\ d\ge1,\\
0, & \quad \text{for}\ w\ge 1,\ d= 0,\\
\frac{\displaystyle { d-1 \choose \lceil\frac{qw}{2} \rceil
    -1 } {qK-d \choose \lfloor \frac{qw}{2} \rfloor } }{\displaystyle {qK \choose qw
  } }, & \quad \text{for}\ w\ge 1,\ d\ge 1.
\end{cases}
\label{eq:Pr_dw0}
\end{equation}
Since we are only interested in code words with nonzero weight,
in the following we will only consider the case where $w\ge 1$. 
Note that from \eqref{eq:Pr_dw0} we obtain the constraints
\begin{equation}
\left\lceil\frac{qw}{2} \right \rceil \leq d \quad\text{and}\quad
\left \lfloor \frac{qw}{2} \right\rfloor \leq qK-d.
\label{eq:lim}
\end{equation}
The total number of input sequences having weight $w$ is ${K \choose
  w}$. 
Then $E(A_{d,w})$ is given as
\begin{equation}
E(A_{d,w}) =  {K \choose  w} \, \Pr(d|w).
\label{eq:Pr_dw}
\end{equation}
By using the fact that
\begin{equation}
N^\ell > \prod_{\lambda=0}^{\ell-1} (N-\lambda) \ge \frac{ N^{\ell} }{\varphi_N(\ell)} \quad
\text{with}\quad \varphi_\lambda(\ell)\triangleq \text{exp}\left(
  \frac{\ell(\ell-1)}{2\lambda} \right),
\label{eq:phi_ell}
\end{equation}
we can show that
\begin{equation}
\left( \frac{N}{\ell}\right)^\ell \frac{1}{\varphi_N(\ell)} \leq {N \choose \ell}
\leq \left( \frac{N}{\ell}\right)^\ell \varphi_\ell(\ell).
\label{eq:bcoeff_bound}
\end{equation}
Thus we can upper bound \eqref{eq:Pr_dw} as
\begin{equation}
  E({A}_{d,w}) \le \frac{N^w\, d^{\lceil qw/2\rceil -1 }\, N^{\lfloor qw/2
    \rfloor}   }{q^w\, N^{qw} }\, 2^{qw}\, \left\lceil \frac{qw}{2}
  \right\rceil\, \varphi_N(w).
\label{eq:Pr_dw1}
\end{equation}
The cumulative WEF is given by
\begin{equation}
  E({A}_{d \le \delta})
  = \sum_{w=1}^{K} \sum_{d=1}^{\delta} E({A}_{d,w}),
\label{eq:sum_dw1}
\end{equation}
and by using \eqref{eq:Pr_dw1} the sum over $d$ in \eqref{eq:sum_dw1}
can be upper bounded as 
\begin{equation}
 \sum_{d=1}^{\delta} E({A}_{d,w})  \le  \sum_{d=1}^{\delta} \frac{N^w\, d^{\lceil qw/2\rceil -1 }\, N^{\lfloor qw/2
    \rfloor}   }{q^w\, N^{qw} }\, 2^{qw}\, \left\lceil \frac{qw}{2}
\right\rceil\, \varphi_N(w) <   \frac{\delta ^{\lceil qw/2\rceil } }{ N^{qw-w-\lfloor qw/2
    \rfloor} }\,\left(\frac{2^{q}}{q}\right)^w\, \left\lceil \frac{qw}{2}
\right\rceil\, \varphi_N(w),
\label{eq:sum1_dw1}
\end{equation}
where $w\ge 1$, $d\ge 1$. Now choosing  $\delta=N^{\frac{q-2}{q}-\epsilon}$, where
$\epsilon$ is any fixed positive value, we obtain
\begin{align}
 \sum_{d=1}^{\delta} E({A}_{d,w})  &<   N^{-qw
 \epsilon/2}\, \left
 (\frac{2^{q}}{q}\right)^w\,\left(e^{(w-1)/(2N)}\right)^w\,\left((qw)^{1/w}\right)^w \notag \\
& <  N^{-qw 
 \epsilon/2}\, \left(\frac{2^{q}}{q}\right)^w\,  \left( e^{1/(2q)}\right)^w \, e^{qw/e}=\eta^w,
\label{eq:sum2_dw1}
\end{align}
where
\begin{equation}
 \eta=  N^{-q \epsilon/2}\, \frac{2^{q}}{q}  \,  e^{1/(2q)} \, e^{q/e},
\label{eq:eta}
\end{equation}
and we have employed  the  inequalities 
\[
\max_{1 \le w \le K} e^{(w-1)/(2N)} < e^{1/(2q)} \quad \text{and}\quad (qw)^{1/w} \le e^{q/e}.
\]
Next we choose an $N_0=N_0(\epsilon)$ such that
\begin{equation}
  \eta_0= N_0 ^{-q
 \epsilon/2}\, \frac{2^{q}}{q} \,  e^{1/2q} \, e^{q/e} < \frac{1}{2},
\label{eq:sum3_dw1}
\end{equation}
and then, for $N \ge N_0$, by combining \eqref{eq:sum_dw1},
\eqref{eq:sum2_dw1}, and \eqref{eq:eta} we obtain
\begin{equation}
  E({A}_{d \le \delta}) < \sum_{w=1}^{\infty} \eta^w =\frac {\eta}{1-\eta}<1.
\label{eq:Mdelta}
\end{equation}
From (\ref{eq:Mdelta}) we conclude that there exists codes in the
ensemble with block length $N>N_0$ whose minimum distance
satisfies $d_{\text{min}}^{\text{RA}}> N^{\frac{q-2}{q}-\epsilon}$. Since
  $\eta/(1-\eta)\rightarrow 0$ as $N \rightarrow
\infty$, the fraction of codes in the ensemble with  $d_{\text{min}}^{\text{RA}}\le
N^{\frac{q-2}{q}-\epsilon}$ goes to zero, which proves the theorem.
\end{proof}

In the case $q=2$ the above bound does not hold. In particular, using
\eqref{eq:Pr_dw} we can show that \linebreak[4] $E(A_{d=1,w=1})=1$,
i.e., on average each RA code has one codeword of weight one generated
by a weight one input sequence.


\section{Ensemble average weight spectrum for
 repeat-accumulate-accumulate codes}
\label{sec:RAA}
\begin{figure*}[htb]
  \centerline{\includegraphics[scale=0.7]{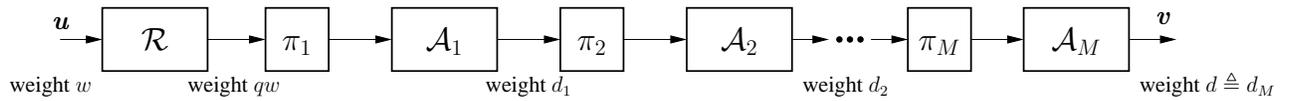}}
    \caption{Repeat multiple accumulate encoder.}
    \label{fig:RAm}
\end{figure*}

In this and the following section, we extend the encoder of Fig.~\ref{fig:RA} and
consider a serial concatenation of $M$  accumulators ${\cal A}_\ell$
with generator polynomials $1/(1+D)$ separated by
interleavers $\pi_\ell$, $1\le \ell \le M$, as shown in
Fig.~\ref{fig:RAm}.
In particular, in this section we focus on $M=2$ and, based on an
average weight enumerator analysis, show that the resulting repeat
double accumulate (RAA) code ensembles for $q\ge 3$ are asymptotically good,
i.e., as the block length $N$ tends to infinity, almost all codes in the
ensemble exhibit linear distance growth with block length.

Analogous to \eqref{eq:Pr_dw0}, the conditional probability that a
weight $d_1$ codeword is obtained at the output of the first
accumulator and a weight $d$ codeword at the output of the second
accumulator for a given input weight $w$ is given as
\begin{equation}
\begin{aligned}
  \Pr(d,d_1|w) =& \Pr(d_1|w) \, \Pr(d|d_1) \\
  =& \frac{\displaystyle  {d_1-1 \choose \lceil\frac{qw}{2} \rceil-1} {qK-d_1
      \choose \lfloor \frac{qw}{2} \rfloor } } { \displaystyle {qK \choose qw} } \,
  \frac{\displaystyle  {d-1 \choose \lceil\frac{d_1}{2} \rceil-1} {qK-d \choose
      \lfloor \frac{d_1}{2} \rfloor } } { \displaystyle {qK \choose d_1} }
\end{aligned}
\label{eq:RAA_prob1}
\end{equation}
where $w$, $d_1$, and $d$ must satisfy  the constraints
\begin{equation}
\left\lceil\frac{qw}{2} \right \rceil \leq d_1, \quad
\left \lfloor \frac{qw}{2} \right\rfloor \leq qK-d_1, \quad
\left\lceil\frac{d_1}{2} \right\rceil \leq d, \quad  \text{and}\quad
\left\lfloor \frac{d_1}{2} \right \rfloor \leq qK-d.
\label{eq:lim1}
\end{equation}
After some straightforward manipulations and, recalling that $N=qK$, \eqref{eq:RAA_prob1} can be rewritten as
\begin{equation}
\Pr(d,d_1|w) 
		= \frac{\displaystyle 
				{ N -qw  \choose d_1-\lceil \frac{qw}{2} \rceil} 
				{qw \choose \lceil \frac{qw}{2} \rceil } 
				{ d_1 \choose \lceil\frac{d_1}{2} \rceil}
				{N-d_1 \choose d-\lceil \frac{d_1}{2} \rceil }  }
				{\displaystyle {N \choose d_1  } \ {N  \choose d   } } 
				\cdot \frac{ \left\lceil \frac{qw}{2} \right\rceil \,
  			\left\lceil \frac{d_1}{2}\right \rceil}{d_1\,d}.
\label{eq:RAA_prob2}
\end{equation}
We can now write the ensemble average conditional IOWEF as 
\begin{equation}
E(A_{d,d_1,w}) = {K \choose w} \, \Pr(d,d_1|w).
\label{eq:RAA_WE}
\end{equation}

An upper bound on the ensemble-average WEF is given by
\begin{equation}
   E({A}_{d}) =\sum_{w=1}^K \sum_{d_1=1}^N E({A}_{d,d_1,w}) 
   \le K\,N \max_{1\le w\le   K} \max_{1\le d_1\le  N} E({A}_{d,d_1,w}),
\label{eq:RAA_WEF}
\end{equation}
and a lower bound is given by
\begin{equation}
     E({A}_{d})  \ge \max_{1\le w\le   K} \max_{1\le d_1\le  N} E({A}_{d,d_1,w}).
\label{eq:RAA_WEF_lb}
\end{equation}
In a similar way, the ensemble-average cumulative WEF $E({A}_{d \le \delta})$ 
can be upper bounded by
\begin{equation}
   E({A}_{d \le \delta})  
   \le \delta \,K\,N \max_{1\le d\le \delta} \max_{1\le w\le   K} 
   \max_{1\le d_1\le  N} E({A}_{d,d_1,w}).
\label{eq:RAA_cWEF}
\end{equation}

In the following, our goal is to show that the ensemble-average 
cumulative WEF tends to zero as $N\rightarrow \infty$ for all $\delta<\hat{\rho}_{\text{min}} N$, where $\hat{\rho}_{\text{min}}$ is
a lower bound on the asymptotic minimum distance growth rate coefficient 
of the ensemble. We do this by showing that 
$N^3 E({A}_{d,d_1,w}) \rightarrow 0$ for all values of $d$,
$1 \le d < \hat{\rho}_{\text{min}}N$.

To this end, we write the weights $w$, $d_1$, and $d$ as
\begin{equation}
   w =  \alpha \, N^a, \qquad
   d_1 =  \beta \, N^b, \qquad
   d =  \rho \, N^c,
\label{eq:cond}
\end{equation}
where $0\leq a\leq b\leq c \leq 1$, and $\alpha$, $\beta$, $\rho$ are
positive constants, and condition \eqref{eq:lim1} must be satisfied.

We now divide the problem into the following two cases:
\begin{enumerate}
	\item 
	At least one of the weights $w$, $d_1$, and $d$ is of order $o(N)$, so 
	at least one of the constants $a$, $b$, and $c$ is less than 1.
	\item
	All the weights $w$, $d_1$, and $d$ can be expressed as fractions of the block length $N$, so $a=b=c=1$.
\end{enumerate}
\begin{lemma}
\label{lem:1}
		In the ensemble of RAA codes with block length $N$ and $q \geq 3$,
		in the case where at least one of the weights $w$, $d_1$, and $d$ is 
		of order $o(N)$, $N^3 E({A}_{d,d_1,w}) \rightarrow 0$ as 
		$N\rightarrow \infty$ for all values of $d$,
		$1 \le d < N/2$.
\end{lemma}
The proof of Lemma \ref{lem:1} can be found in Appendix \ref{sec:proof_lemma1}. 
From Lemma \ref{lem:1} we conclude that the contribution of the first case to
the cumulative WEF tends to zero as the block length $N$ tend to infinity.
Thus it is sufficient to only consider weights that are of the same order as the
block length $N$.

We now consider the second case when all the weights
can be expressed as fractions of the block length $N$.
Using Stirling's approximation, we have that 
${N \choose d}\approx e^{N\cdot H(d/N)}$, where
$H(x)=-x\,\ln(x)-(1-x)\,\ln(1-x)$ is the binary entropy
function. Applying this to \eqref{eq:RAA_WE}, 
for $N \rightarrow \infty$ we obtain
\begin{equation}
 E({A}_{d,d_1,w}) =\exp \left( f(\alpha, \beta, \rho)\,N+o(N)\right ), 
\label{eq:RAA_exp}
\end{equation}
where $\alpha=w/K=qw/N$, $\beta=d_1/N$, and $\rho=d/N$ are normalized
weights, and the function
\begin{multline}
 f(\alpha, \beta, \rho)= \frac{H(\alpha)}{q} -H(\beta) -H(\rho) +
 H\left(\frac{\beta -\alpha /2}{1-\alpha}\right)(1-\alpha)\\ +\alpha\, \ln 2 +
 H\left(\frac{\rho -\beta /2}{1-\beta} \right )(1-\beta ) +\beta\,\ln 2.
\label{eq:RAA_f}
\end{multline}

\begin{figure}[htb]
  \centerline{\includegraphics[scale=1.1]{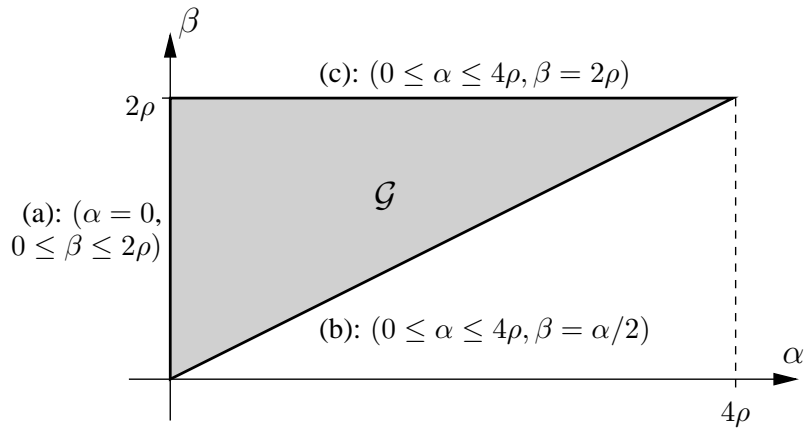}}
    \caption{Properties of the function $f(\alpha,\beta,\rho)$:
      region ${\cal G}$ for $\rho<0.25$.}
    \label{fig:region}
\end{figure}

Since the minimum distance of any linear code cannot be greater than $N/2$,
we consider a fixed $\rho\le 1/2$, and the function $f(\alpha,\beta,\rho)$ in
\eqref{eq:RAA_f} is defined on a region ${\cal G}\subset \Rset^2$ with
boundaries
\begin{equation}
\begin{split}
\text{(a)\quad}&(\alpha =0, 0 \leq \beta \leq
2\rho),\\
\text{(b)\quad}&(0 < \alpha \le
\min(1,4\rho), \beta =\alpha /2),\\
\text{(c)\quad}&(0 < \alpha \leq
\min(2-4\rho,4\rho), \beta = 2\rho),\\
\text{(d)\quad}& (2\,(1-2\rho)\leq \alpha \leq 1 ,
\beta =1-\alpha/2)\quad \text{if}\quad\rho>0.25,\\
\end{split}
\label{eq:boundaries}
\end{equation}
where the boundaries follow from the constraints in \eqref{eq:lim1}.
Note that boundary (a) should be excluded from
region ${\cal G}$ since we exclude the case $w=0$. However,
for the sake of clarity, we call it a boundary of ${\cal G}$ since
all points $(\alpha>0,\beta>0)$ arbitrarily close to boundary (a) in
${\cal G}$ must be included. An example of ${\cal G}$ is shown for $\rho<0.25$
in Fig.~\ref{fig:region}.
Now let $\tilde{f}_{\rho}(\alpha,\beta)\triangleq f(\alpha,\beta,\rho)$ be
a function of $\alpha$ and $\beta$ for a fixed $\rho<1/2$.
From (\ref{eq:RAA_f}) it can be seen that the value of the function
$\tilde{f}_\rho(0,0)$ is zero for any given $\rho$. 
A sufficient condition for linear distance growth can be stated as
follows: if $\tilde{f}_\rho(\alpha,\beta)$ is strictly negative for
all $0<\rho< \hat{\rho}_{\text{min}}$ and all $\alpha$, $0< \alpha \le
2\min(\beta,1-\beta)$, and $\beta$, $0< \beta \le 2\min(\rho,
1-\rho)$, then almost all codes in the ensemble have minimum distance
$d_{\text{min}}=\rho_{\text{min}}\,N\ge \hat{\rho}_{\text{min}}\,N$.
Thus, to show that $\hat{\rho}_{\text{min}}$ is a lower bound on the 
asymptotic minimum distance growth rate coefficient of the ensemble 
the maximum of $\tilde{f}_\rho(\alpha,\beta)$ for all
$0<\rho<\hat{\rho}_{\text{min}}$  must be negative in ${\cal G}$.

We now address the maximization of $f(\alpha, \beta, \rho)$ over
$\alpha$ and $\beta$, where, in principle, a maximum of the function
$\tilde{f}_\rho(\alpha,\beta)$ can occur inside the region ${\cal G}$
or on boundaries (b), (c), or (d). However, we will show below that
the maximum only occurs inside the region ${\cal G}$.

\begin{lemma}
\label{lem:2}
For any given $\rho\le 1/2$, the stationary points of the function
$\tilde{f}_\rho(\alpha,\beta)$ satisfy the following system of
equations:
\begin{gather}
 \left(\frac {\beta -\alpha /2}{1-\alpha}\right)\, \left(\frac{1-\beta
     -\alpha /2}{1-\alpha}\right )=\left (\frac{\alpha}{1-\alpha}\right)^{\frac {2}{q}},
\label{eq:stp}\\
4\, \left(\frac {\rho -\beta /2}{1-\beta}\right)\, \left(\frac{1-\rho -\beta
  /2}{1-\beta}\right)=\left(\frac{1-\beta}{\beta} \, \frac{\beta -\alpha /2}{1-
  \beta -\alpha /2}\right)^{2}.
\label{eq:str1}
\end{gather}
\end{lemma}

\begin{proof}
The partial derivatives of the function  $\tilde{f}_{\rho}(\alpha,\beta)$
are given as 
\begin{gather}
\frac{\partial \tilde{f}}{\partial \alpha}= -\frac{1}{q}\, \ln \left(\frac{\alpha}{1-\alpha} \right)+
\frac{1}{2}\,\ln  \left(\frac{\beta -\alpha /2}{1-\alpha}\right)+ \frac{1}{2}\,\ln
 \left(\frac{1-\beta -\alpha /2}{1-\alpha}\right)+\ln 2, \label{eq:stpal}\\
\frac{\partial \tilde{f}}{\partial \beta} =  \ln \left( \frac{\beta}{1-\beta}\right)- \ln  \left(\frac{\beta -\alpha /2}{1-
  \beta -\alpha /2}\right) +
\frac{1}{2}\,\ln  \left( \frac{\rho -\beta /2}{1-\beta}\right)+ \frac{1}{2}\,\ln
\left(\frac{1-\rho -\beta /2}{1-\beta}\right)+\ln 2.
\label{eq:stpbe}
\end{gather}
At the stationary points of $\tilde{f}_{\rho}(\alpha,\beta)$, we have
$\partial \tilde{f}/ \partial\alpha=0$ and $\partial \tilde{f}/
\partial\beta=0$. By setting  \eqref{eq:stpal} to zero and letting 
$x\triangleq \frac{\beta-\alpha/2}{1-\alpha}$, we obtain 
\[
4\,x\,(1-x)=\left(\frac{\alpha}{1-\alpha}\right)^{\frac{2}{q}},
\]
which is identical to \eqref{eq:stp}. Likewise, by setting
\eqref{eq:stpbe} to zero 
with $y\triangleq \frac{\rho-\beta/2}{1-\beta}$ we obtain
\[
4\,y^2-4\,y+\left( \frac{1-\beta}{\beta}\right)^2\,
\left(\frac{\beta-\alpha/2}{1-\beta-\alpha/2} \right)^2 =0,
\]
which leads to \eqref{eq:str1}.
\end{proof}

In order to find the stationary points of $\tilde{f}_\rho(\alpha,\beta)$, 
we must solve \eqref{eq:stp} and \eqref{eq:str1}, where $\rho$ is expressed
as a function of $\alpha$ and $\beta$.  We first consider (\ref{eq:str1}),
which can be viewed as a quadratic equation in $\rho$ with variables 
$\alpha$ and $\beta$. It follows that, if $\beta> 1/2$, (\ref{eq:str1}) 
has no real-valued zeros.  On the other hand, if $\beta \leq 1/2$, then
(\ref{eq:str1}) contains two real-valued zeros: $\rho_1(\alpha,\beta)
\leq 1/2$ and $\rho_2(\alpha,\beta) \geq 1/2$. Since we
only consider the case $\rho\le 1/2$, the solution of the
quadratic equation \eqref{eq:str1} in $\rho$ then yields
\begin{equation}
\rho=\rho_1 (\alpha, \beta)= \frac {1}{2} -  \frac {1-\beta}{2}
\sqrt{1-\left(\frac{1-\beta}{\beta} \, \frac{\beta -\alpha /2}{1-\beta -\alpha /2}\right)^{2}},
\label{eq:root1}
\end{equation}
which relates the quantities $\alpha$, $\beta$, and $\rho$.

Now consider (\ref{eq:stp}) as a quadratic equation in $\beta$ with
variable $\alpha$. For
$\alpha>1/2$ there are no real-valued zeros, but for $\alpha \leq 1/2$
\eqref{eq:stp} has two real-valued zeros: $\beta_1(\alpha) \leq 1/2$
and $\beta_2(\alpha) \geq 1/2$. Since $\rho_1(\alpha,\beta)$ is
real-valued only if $\beta \leq 1/2$, the solution of the quadratic
equation (\ref{eq:stp}) in  $\beta$ is given by
\begin{equation}
\beta=\beta_1(\alpha)= \frac {1}{2} -  \frac
{1-\alpha}{2}\sqrt{1-\left(\frac{\alpha}{1-\alpha}\right)^{\frac {2}{q}}}.
\label{eq:sol_beta}
\end{equation}

A pair $(\alpha,\beta)$ maximizing\footnote{The structure
  of (\ref{eq:root1}) and (\ref{eq:sol_beta}) suggest a slightly
  different but equivalent approach, where $\alpha$ is used as the
  free parameter and $\beta$ and $\rho$ are determined from
  (\ref{eq:root1}) and (\ref{eq:sol_beta}). } 
$\tilde{f}_\rho(\alpha,\beta)$ can be computed from (\ref{eq:root1}) and
(\ref{eq:sol_beta}) for a given $\rho$. 
There is a lower limit $\rho_0$ for which the system of equations has
a solution.  If $\rho<\rho_{0}$, (\ref{eq:root1}) and
(\ref{eq:sol_beta}) have no solution and
$\tilde{f}_{\rho}(\alpha,\beta)$ has no stationary points within
${\cal G}$.
For $\rho>\rho_0$, (\ref{eq:root1}) and (\ref{eq:sol_beta}) yield 
two solutions, where one solution corresponds to a maximum and the 
other to a saddle point. 
Finally, there is only a single stationary point for $\rho=\rho_0$.
\begin{figure}[htb]
  \centerline{\includegraphics[width=0.65\textwidth]{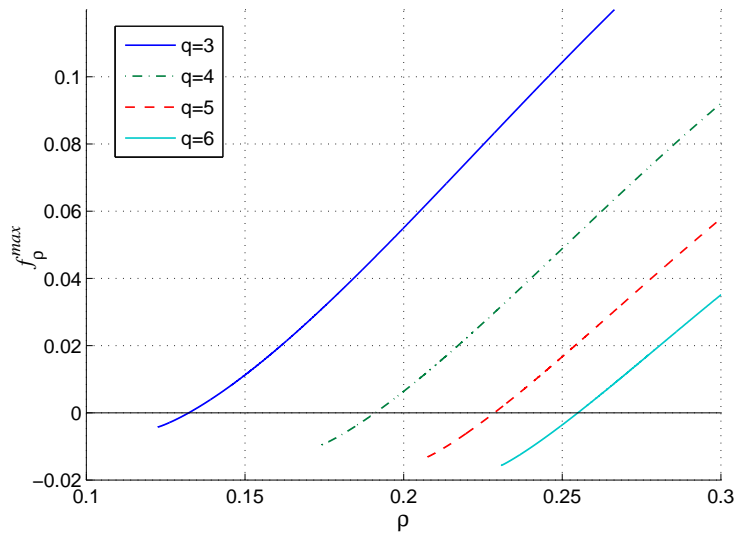}}
  \caption{Maxima $\tilde{f}^\text{max}_\rho$ inside the region ${\cal G}$
    for the RAA code ensemble with rates
    $R=1/3,1/4,1/5,1/6$.} 
    \label{fig:RAA} 
\end{figure}
Fig.~\ref{fig:RAA} shows the maxima $\tilde{f}^\text{max}_\rho$ inside
the region ${\cal G}$ for different values of $q$, where we observe a
zero crossing at $\rho=\hat{\rho}_{\text{min}}$,
$0<\hat{\rho}_{\text{min}}\le 0.5$. 
To show that $\hat{\rho}_{\text{min}}$ is indeed the asymptotic minimum
distance growth rate coefficient, we now address the behavior of the function
$\tilde{f}_{\rho}(\alpha,\beta)$ at the  boundaries of ${\cal G}$.

\begin{lemma}
\label{lem:3}
The function $\tilde{f}_{\rho}(\alpha,\beta)$ cannot have a maximum 
on boundaries (a), (b), and (d) but can have a maximum on
boundary (c).
\end{lemma}

\begin{figure}[htb]
  \centerline{
  \subfigure[$\rho=0.1$]{\includegraphics[scale=0.45]{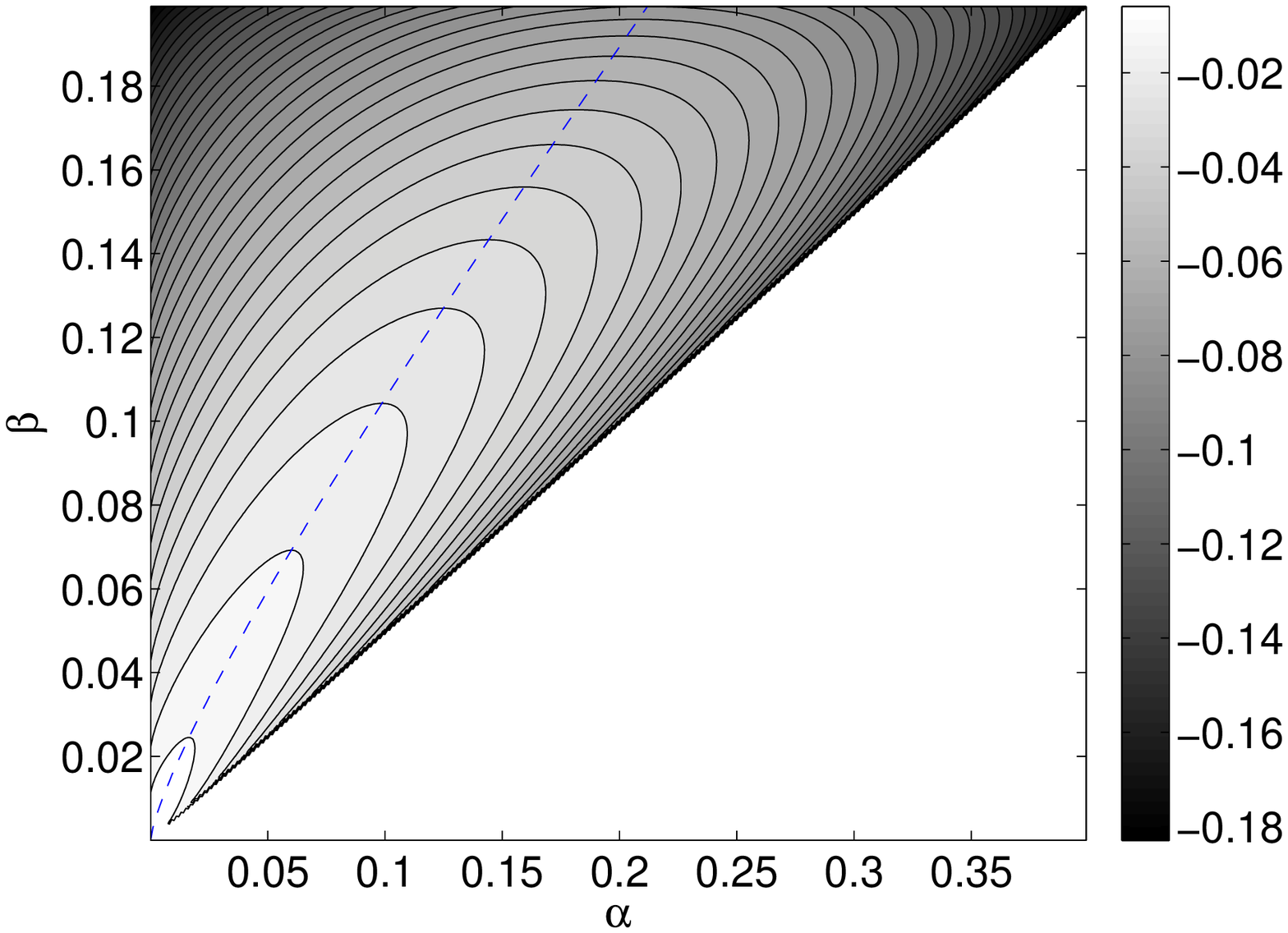}}
  \subfigure[$\rho=0.2$]{\includegraphics[scale=0.45]{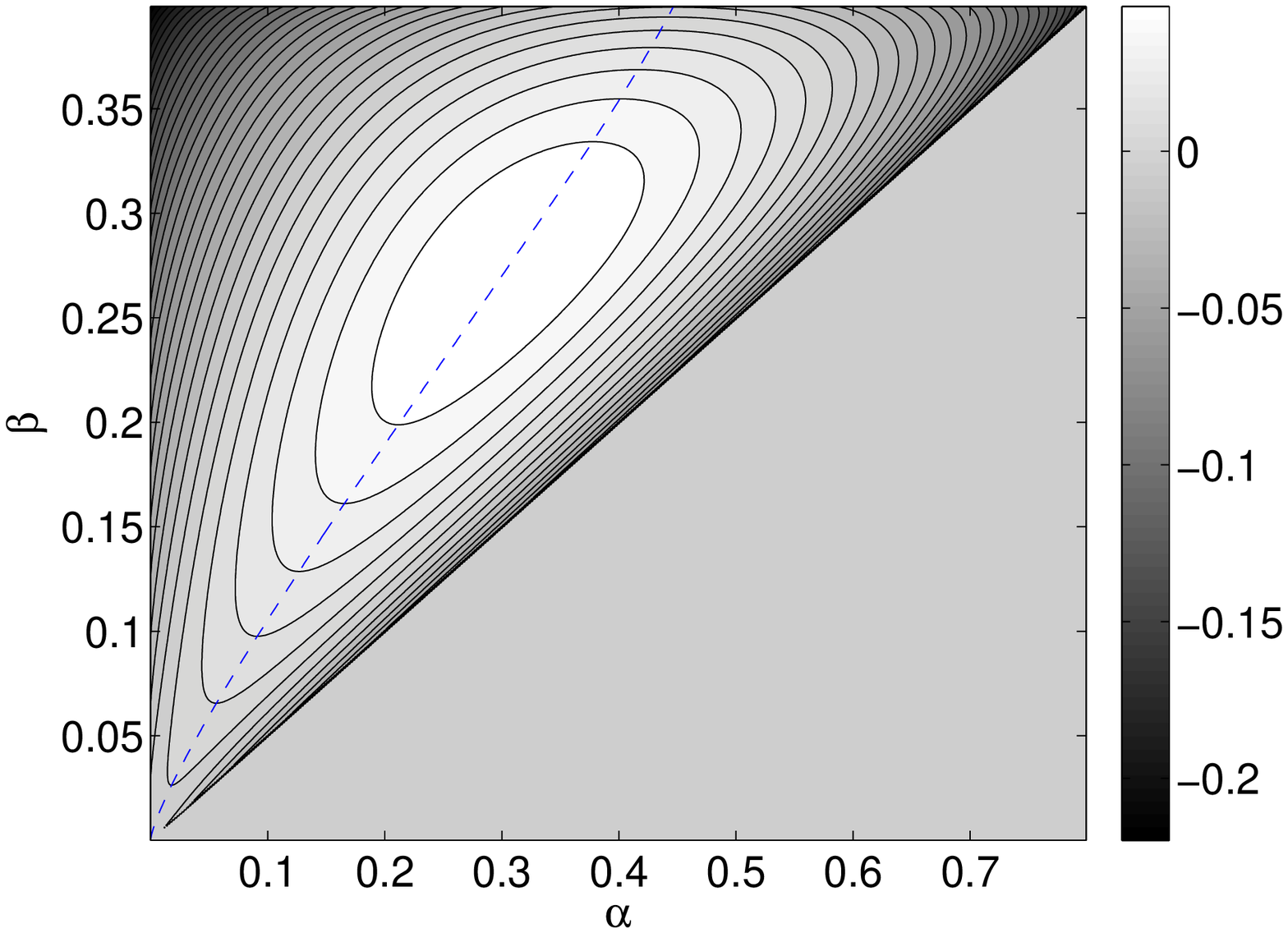}}}
    \caption{Contour plots of
      $\tilde{f}_{\rho}(\alpha,\beta)=f(\alpha,\beta,\rho)$ for a fixed $\rho$ over the region
      ${\cal G}$. Maximum values can only be located along the
      dashed 
      line within ${\cal G}$.}
    \label{fig:f_contour}
\end{figure}

\begin{proof}
	We introduce the function $\varphi_{\rho,\beta}(\alpha)\triangleq
f(\alpha,\beta,\rho)$ for  fixed $\beta$ and $\rho$, which is 
defined on a line parallel to the $\alpha$-axis
(see Fig.~\ref{fig:f_contour} for an illustration), and consider the
following three cases:
\begin{enumerate}
\item For $0< \beta \le
  \min(0.5,2\rho)$, we fix the normalized weight $\beta$ and allow
  $\alpha$ to vary between $0$ and
  $2\beta$. In this case, $\varphi_{\rho,\beta}(\alpha)$ is defined on a line
  from boundary~(a) to (b) in \eqref{eq:boundaries}.
\item For $1/2< \beta \le 2\rho$, we fix $\beta$ and allow  $\alpha$
  to vary between $0$ and
  $2(1-\beta)$. In this case, the function $\varphi_{\rho,\beta}(\alpha)$ is
  defined on a line from boundary (a) to (d).
\item For $\beta=2\rho$,  the normalized input weight $\alpha$ varies between $0$ and
  $\min(2-4\rho,4\rho)$, i.e.,  $\varphi_{\rho,\beta}(\alpha)$ is
  defined on boundary (c).
\end{enumerate} 
The derivative $d\varphi_{\rho,\beta}(\alpha)/d\alpha$ is given by
\eqref{eq:stpal} in all three cases. Note that, at the stationary
points, $\alpha$ and $\beta$ are related by \eqref{eq:stpal},
independent of $\rho$: if $\beta\le 1/2$, then $\alpha$ can be
obtained by solving \eqref{eq:sol_beta}, and if $\beta> 1/2$, we can
find $\alpha$ by solving
\[
\beta= \frac {1}{2} +  \frac
{1-\alpha}{2}\sqrt{1-\left(\frac{\alpha}{1-\alpha}\right)^{\frac {2}{q}}}.
\]
The second derivative $d^2 \varphi_{\rho,\beta}(\alpha)/d\alpha^2$ is
given by
\begin{equation}
\frac{d^2
  \varphi_{\rho,\beta}(\alpha)}{d\alpha^2}=-\frac{1}{q}\frac{1}{\alpha\,(1-\alpha)}+
\frac{1}{2}\left( \frac{1-2\,x}{x\,(1-x)}\right)
\frac{2\,\beta-1}{(1-\alpha)^2},
\label{eq:d2alpha}
\end{equation}
where $x\triangleq
\frac{\beta-\alpha/2}{1-\alpha}$ and  
$0\le x<1/2$. At a stationary point with $\alpha=\alpha_0$, the corresponding $\beta$
can be obtained from \eqref{eq:sol_beta}. For such a  pair 
$(\alpha_0,\beta)$, we then obtain from \eqref{eq:d2alpha} that
\[
\frac{d^2
  \varphi_{\rho,\beta}(\alpha)}{d\alpha^2}\biggr|_{\alpha=\alpha_0}=
-\frac{1}{q\,\alpha_0\,(1-\alpha_0)} -
\frac{(1-2\,\beta)^2}{(1-\alpha_0)\,(\beta-\alpha_0/2)\,(1-\beta-\alpha_0/2)}
<0.
\]
It follows that the stationary point on a constant $\beta$ line in the $(\alpha,\beta)$
plane corresponds to a maximum for each $(\alpha_0,\beta)$ pair
satisfying \eqref{eq:sol_beta}. Consequently, for cases 1) and 2), the
maximum of the function $\tilde{f}_\rho(\alpha,\beta)$ cannot be on
the boundaries (a), (b), and (d). For the same reason, in  case 3), the maximum of
the line $\tilde{f}_\rho(\alpha,\beta)\bigr|_{\beta=2\rho}$ is located on the boundary
(c).
\end{proof}

\begin{lemma}
\label{lem:4}
For any
$\rho<1/2$ and $q \geq 3$ the function $f(\alpha,\beta,\rho)$ is negative on the
boundaries (a) and (b) of the region ${\cal G}$ except for the
point $(\alpha,\beta)=(0,0)$, where $f(0,0,\rho)=0$ for any
$\rho<1/2$.
\end{lemma}

The proof of Lemma \ref{lem:4} can be found in Appendix~\ref{sec:proof_lemma4}.

A numerical analysis shows that the maximum value on boundary (c) is
always less than the maximum inside the region ${\cal G}$, if it
exists, or strictly negative if there is no stationary point inside
${\cal G}$. And since the function $f(\alpha,\beta,\rho)$ is always 
negative on the boundaries (a) and (b), except for the point 
$(\alpha,\beta)=(0,0)$, we need not consider the values on the boundary of
the region ${\cal G}$ in \eqref{eq:boundaries}, and we conclude that,
for all $\rho<\hat{\rho}_{\text{min}}$, the function
$\tilde{f}_\rho(\alpha,\beta)$ is negative. Thus, we obtain from
Lemmas~\ref{lem:2}, \ref{lem:3}, and \ref{lem:4} that 
$\hat{\rho}_{\text{min}}$ is a lower bound on the asymptotic minimum
distance growth rate coefficient of the code ensemble.

We summarize our findings in the following theorem.
\begin{theorem}
\label{th:RAA}
  In the ensemble of RAA codes of rate $R\le 1/3$ and block length
  $N\rightarrow\infty$, almost all codes have minimum distance
  $d_{\text{min}}$ growing linearly with $N$. A lower bound on the
  asymptotic minimum distance growth rate coefficient
  $\rho^{RAA}_{\text{min}}=d^{RAA}_{\text{min}}/N$ of
  the ensemble can be obtained by solving the system of equations
  (\ref{eq:root1}) and (\ref{eq:sol_beta}), i.e., by finding the maximum
  of the function $f(\alpha,\beta,\rho)$.
\end{theorem}

To illustrate the behavior of the function $\tilde{f}_{\rho}(\alpha,\beta)$,
Fig.~\ref{fig:f_contour} shows two examples of contour plots of $\tilde{f}_{\rho}(\alpha,\beta)$
for the RAA code ensemble with $q=3$ and different values of $\rho$.
For $\rho=0.1$, displayed in Fig.~\ref{fig:f_contour}(a), there is no
stationary point inside the region ${\cal G}$.
The function $\tilde{f}_{\rho}(\alpha,\beta)$ is decreasing monotonically 
from the origin towards boundary (c), located at the top of 
Fig.~\ref{fig:f_contour}(a).
By contrast, for $\rho=0.2$, Fig.~\ref{fig:f_contour}(b) clearly shows a maximum
inside ${\cal G}$. Note that the pairs $(\alpha_0,\beta)$ satisfying \eqref{eq:sol_beta}
correspond to the dashed lines in Fig.~\ref{fig:f_contour}, which
indicate the possible locations of stationary points in the
$(\alpha,\beta)$ plane.

The values of $\hat{\rho}_{\text{min}}$ are listed 
in Table~\ref{tab:bound} along with the corresponding
values of the GVB, the estimated minimum distance growth rate coefficient
$\tilde{\rho}_{\text{min}}$ based on a finite block length analysis
analogous to the one presented in Fig.~\ref{fig:finite}, and the values
$\rho_0$.  We see that, especially for lower code rates, the asymptotic
minimum distance growth rate coefficient of the RAA ensemble is close to
the GVB. Also, the results of the finite length analysis 
match the asymptotic results quite well. 

\begin{table}[htb]
  \caption{Asymptotic minimum distance growth rate coefficient 
    lower bound $\hat{\rho}_{\text{min}}$, estimated
    minimum distance growth rate coefficient
    $\tilde{\rho}_{\text{min}}$  from a
    finite length analysis, and the corresponding
    values of the GVB for the RAA ensemble
    with different code rates. The values $\rho_0$
    denote the smallest $\rho$ for which a solution of the
    maximization problem given by \eqref{eq:stp} and \eqref{eq:str1} can be obtained.}
\label{tab:bound}
\begin{center}
\begin{tabular}{|c|c|c|c|c|}
\hline
$q$\st  & $\hat{\rho}_{\text{min}}$ &$\tilde{\rho}_{\text{min}}$ & GVB  &
$\rho_0$
\\\hline \hline
3 \st & 0.1323 & 0.1339 & 0.1740 & 0.1225 \\\hline
4 \st & 0.1911 & 0.1935 & 0.2145 & 0.1742 \\\hline
5 \st & 0.2286 & 0.2312 & 0.2430 & 0.2075 \\\hline
6 \st & 0.2549 & 0.2570 & 0.2644 & 0.2309 \\
\hline
\end{tabular}
\end{center}
\end{table}

For $q=2$, there exists no such lower bound on the asymptotic minimum
distance growth rate coefficient. In this case, for any $\bar{\rho}>0$
the cumulative WEF of the RAA code ensemble can be lower bounded by
\begin{equation}
\begin{aligned}
	E({A}_{d \le \bar{\rho} N})
  = &  \sum_{w=1}^{K} \sum_{d_1=1}^{N} \sum_{d=1}^{\bar{\rho} N} E({A}_{d,d_1,w}) \\
  \geq & \sum_{d=1}^{\bar{\rho} N} E({A}_{d,d_1=1,w=1}) = \sum_{d=1}^{\bar{\rho} N} \frac{1}{N}
  	= \bar{\rho}.
\end{aligned}
\end{equation}
Even though we expect the majority of codes in the ensemble to have a
minimum distance that grows linearly with block length \cite{Pfi03},
for any fixed $\bar{\rho}>0$, there is a nonvanishing fraction of codes in the
ensemble with minimum distance $d_{\text{min}}<\bar{\rho} N$. Thus, for the
RAA ensemble with $q=2$, we cannot give a lower bound on the
asymptotic minimum distance growth rate coefficient.

\section{Ensemble Average Weight  Spectrum for Repeat Multiple
  Accumulate Codes}

We now generalize the results of the previous section to RMA codes
with $M>2$, i.e., with more than two accumulators (see
Fig.~\ref{fig:RAm})\footnote{In contrast to the $M=2$ case, for $M>2$
  we are able to derive a lower bound on the asymptotic minimum
  distance growth rate coefficient for the rate $R=1/2$ code
  ensemble.}.  In this case, the conditional probability of the weight
vector $\mathbf{d}=[d_1,d_2,\dots,d_M]$ for a given input weight can
be written as
\begin{equation}
\Pr(\mathbf{d}|w)=\Pr(d_1|w)\cdot \prod_{\ell=2}^M \frac{\displaystyle
  { d_\ell -1 \choose \lceil\frac{d_{\ell-1}}{2} \rceil
    -1 } {qK-d_\ell \choose \lfloor \frac{d_{\ell-1}}{2} \rfloor }
}{\displaystyle {qK \choose d_{\ell-1} }},
\label{eq:Pr_d_w}
\end{equation}
where $\Pr(d_1|w)$ is defined in (\ref{eq:Pr_dw0}). The
ensemble average  IOWEF is then given by 
\begin{equation}
E({A}_{\mathbf{d},w}) =  {K \choose  w} \, \Pr(\mathbf{d}|w)=\exp \left(
  f(\boldsymbol{\gamma})\,N+o(N)\right ),
\label{eq:IOWEF_Gen}
\end{equation}
where Stirling's approximation for large $N$ has again been employed.
The vector $\boldsymbol\gamma$ contains normalized weights and is
given by
\[
\boldsymbol\gamma=[\beta_0,\beta_1,\beta_2,\dots,\beta_M]=\left[\alpha=\frac{w}{K},
  \frac{d_1}{N}, \frac{d_2}{N}, \dots,\rho=\frac{d_M}{N}\right],
\]
where $\beta_0\triangleq \alpha$ and $\beta_M\triangleq \rho$. 
The  function $f(\boldsymbol{\gamma})$ in (\ref{eq:IOWEF_Gen}) can now
be written as
\begin{equation}
f(\boldsymbol\gamma)=\frac{H(\alpha)}{q} -\sum_{\ell=1}^{M-1}  H(\beta_\ell)\\
+ \sum_{\ell=1}^M H\left(
  \frac{\beta_\ell-\beta_{\ell-1}/2}{1-\beta_{\ell-1}}\right)\,
(1-\beta_{\ell-1})+\ln 2\,\sum_{\ell=0}^{M-1}\beta_\ell,
\label{eq:f_gen}
\end{equation}
which represents a generalization of the function defined in
\eqref{eq:RAA_f} to more than three normalized weight terms.
Analogous to the derivation for the RAA case in Section~\ref{sec:RAA},
$f(\boldsymbol\gamma)$ must now be maximized over
$\alpha,\beta_1,\dots,\beta_{M-1}$.
Here, the same arguments for the
existence of stationary points on the boundary or inside an
$M$-dimensional region ${\cal G}_M$ can be made, analogous to the RAA
  case, where again only the maximum inside ${\cal G}_M$ must
    be considered in the maximization problem. Thus, the $M+1$ tuple
    $(\alpha,\beta_1,\dots,\beta_{M-1},\rho)$ maximizing
    $f(\boldsymbol\gamma)$ can be expressed by the following
    set of recursive equations:
\begin{gather}
\beta_1=\frac {1}{2} -  \frac {1-\alpha}{2}\sqrt
{1-\left(\frac{\alpha}{1-\alpha}\right )^{\frac {2}{q}}} \quad\text{and}
\label{eq:gamma_1}\\
\beta_{\ell+1}=\frac {1}{2} -  \frac {1-\beta_\ell}{2}\sqrt
  {1-\left (\frac{1-\beta_\ell}{\beta_\ell} \cdot  \frac{\beta_{\ell} -\beta_{\ell-1} /2}{1-
      \beta_\ell -\beta_{\ell-1} /2}\right)^{2}},
\label{eq:gamma_l}
\end{gather}
$1\le \ell\le M-1$, for any $\alpha$ such that $0<\alpha\le 1/2$. 
The derivation of this set of equations
follows from the derivation of \eqref{eq:root1} and \eqref{eq:sol_beta}
in a straightforward way: \eqref{eq:gamma_1} is equivalent to
\eqref{eq:sol_beta} with $\beta$ replaced by $\beta_1$, and
\eqref{eq:gamma_l} is a generalization of  \eqref{eq:root1} with
$\rho$ replaced by $\beta_{\ell+1}$ and $\beta$ by
$\beta_\ell$.

\begin{figure}[htb]
  \centerline{\includegraphics[width=0.7\textwidth]{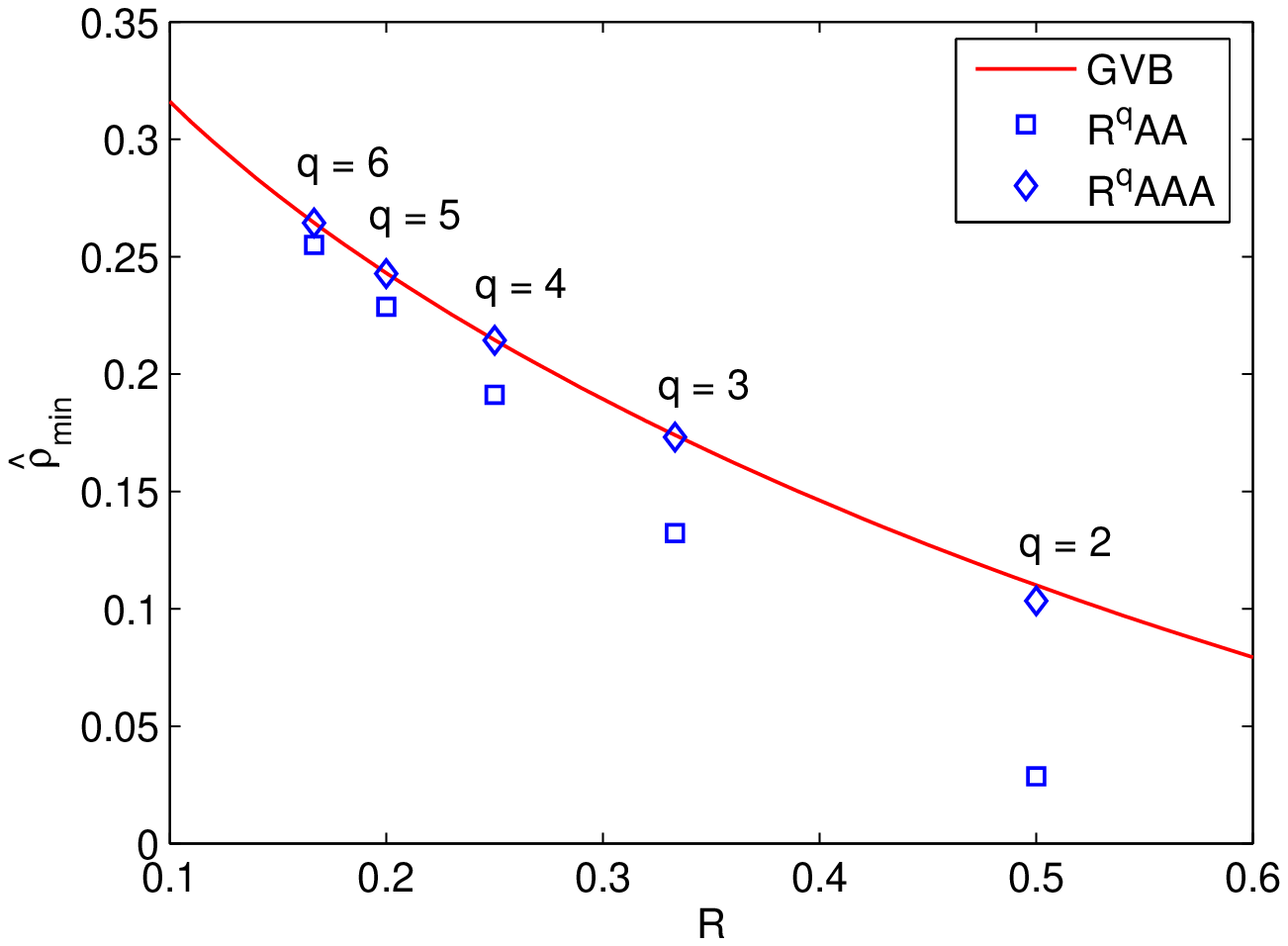}}
  \caption{GVB and the corresponding asymptotic minimum distance growth
  	rate coefficient lower bound $\hat{\rho}_{\text{min}}$ for RAA and RAAA code
    ensembles with rates $R=1/2,1/3,1/4,1/5,\text{ and }1/6$.}
    \label{fig:RAAA}
\end{figure}

As an example,  in Fig.~\ref{fig:RAAA} the values $\hat{\rho}_{\text{min}}$
for the asymptotic minimum distance growth rate coefficient are shown 
for $M=2$ and $M=3$ code ensembles for $q=2,3,4,5,\text{ and }6$
and compared to the GVB.
The values of $\hat{\rho}_{\text{min}}$, along with the estimated values
$\tilde{\rho}_{\text{min}}$ obtained from a finite-length analysis and the
GVB, are listed in Table~\ref{tab:bound_RAAA}.
\begin{table}[htb]
  \caption{Asymptotic minimum distance growth rate coefficient 
    lower bound $\hat{\rho}_{\text{min}}$, estimated minimum distance
    growth rate coefficient $\tilde{\rho}_{\text{min}}$ from a
    finite length analysis, and the corresponding
    values of the GVB for the RAAA ensemble with different code rates.}
\label{tab:bound_RAAA}
\begin{center}
\begin{tabular}{|c|c|c|c|}
\hline
$q$\st  & $\hat{\rho}_{\text{min}}$ & $\tilde{\rho}_{\text{min}}$ & GVB \\\hline \hline
2 \st & 0.1034 & 0.1109 & 0.1100   \\\hline
3 \st & 0.1731 & 0.1739  &0.1740 \\\hline
4 \st & 0.2143 & 0.2150  &0.2145 \\\hline
5 \st & 0.2428 & 0.2400  &0.2430 \\\hline
6 \st & 0.2643 & 0.2627  &0.2644 \\
\hline
\end{tabular}
\end{center}
\end{table}
We observe that the resulting asymptotic minimum distance growth rate coefficients
of the RAAA code ensemble for $q\ge 3$ essentially achieve the GVB,
which is consistent with the results obtained in \cite{PS03} for
finite block lengths and a number of accumulators tending to infinity.

Analogous to Theorem~\ref{th:RAA}, we now state the following
theorem.
\begin{theorem}
\label{th:MRAA}
  In the ensemble of RMA codes with $M$ accumulators, $M\ge 3$, 
  of rate $R\le 1/2$ and  block length $N\rightarrow\infty$, 
  almost all codes have minimum distance $d_{\text{min}}$ growing 
  linearly with $N$. A lower bound on the asymptotic minimum distance
  growth rate coefficient $\rho^\text{RMA}_{\text{min}}=d^\text{RMA}_{\text{min}}/N$ of
  the ensemble can be obtained by solving the system of equations 
  \eqref{eq:gamma_1} and \eqref{eq:gamma_l}, i.e., by finding the 
  maximum of the function $f(\boldsymbol{\gamma})$.
\end{theorem}

\section{Repeat multiple  accumulate codes with random puncturing}
\label{sec:punct}

The rate of RMA code ensembles is determined by the rate of the outer 
repetition code. Thus it is not possible to obtain rates higher than 1/2 
without puncturing. This motivates us to employ random puncturing at 
the output of the inner accumulator in connection with a lower-rate 
RMA mother code for the purpose of achieving higher rate RMA code ensembles.

Let $N'$ be the number of code symbols after puncturing, $d'$ the
corresponding codeword weight, and $R'=R\cdot N/N'$ the code rate.
We also define the ratios $\eta=N'/N$, the normalized block length,  and $\rho'=d'/N'$, the
normalized output weight, after puncturing. The conditional probability
of a weight-$d'$ sequence after puncturing is given by the
hypergeometric distribution
\[
\Pr\mbox{}_{N'}(d'|d)=\frac{\displaystyle { d \choose d' } {N-d \choose N'-d' }
}{\displaystyle { N \choose N'} },
\]
which for large $N$ can be expressed as
\begin{equation}
\Pr\mbox{}_{N'}(d'|d)=\exp\left \{ N \left[ H\left(  \frac{\eta\,\rho'}{\rho}\right) \rho
  + \right. \right. 
\left. \left. H \left(  \frac{\eta(1-\rho')}{1-\rho}\right)(1-\rho)  -
    H(\eta) \right]  +o(N) \right\}.
\end{equation}

Considering the general case of repeat multiple accumulate codes, the
ensemble average  IOWEF can  now be obtained from (\ref{eq:IOWEF_Gen})
as follows:
\begin{align*}
E({A}_{\mathbf{d},d',w}) &=  {K \choose  w} \, \Pr(\mathbf{d}|w)\,
\Pr\mbox{}_{N'}(d'|d) \\
&=\exp \left(  F(\boldsymbol{\gamma},\rho',\eta)\,N+O(\ln N)\right ),
\end{align*}
where $\Pr(\mathbf{d}|w)$ is defined in \eqref{eq:Pr_d_w}, 
$F(\boldsymbol{\gamma},\rho',\eta)\triangleq f(\boldsymbol{\gamma})+\varphi(\rho',\rho,\eta)$,
and 
\begin{equation}
\varphi(\rho',\rho,\eta)=H\left(  \frac{\eta\,\rho'}{\rho}\right) \rho
  + H \left(  \frac{\eta(1-\rho')}{1-\rho}\right)(1-\rho)  - H(\eta).
\label{eq:varphi}
\end{equation}
Following the approach used for the RAA ensemble in Section~\ref{sec:RAA}, 
the maximization of the function $F(\boldsymbol{\gamma},\rho',\eta)$
must now be carried out over all elements of the vector $\boldsymbol{\gamma}$,
including $\rho=\beta_M$. Again, for the same reasons as in the RAA case, 
we consider only stationary points of $F(\boldsymbol{\gamma},\rho',\eta)$
inside the  $M+1$-dimensional region spanned by the $M+1$ tuple
$\boldsymbol{\gamma}$.

Note that $\varphi(\rho',\rho,\eta)$ in (\ref{eq:varphi}) does not depend
on the variables $\alpha,\beta_1,\dots,\beta_{M-1}$, since only the
output of the inner encoder is punctured. Therefore we can still make
use of (\ref{eq:gamma_1}) and (\ref{eq:gamma_l}) for $1\le \ell \le
M-1$. In addition, we need to compute the derivative $\partial
F/\partial\rho$, which is given by
\begin{equation}
\frac{\partial F}{\partial\rho}=\ln\left(
  \frac{\rho^2\,(1-\rho-\beta_{M-1}/2)}{(1-\rho)^2\,
    (\rho-\beta_{M-1}/2)}   \right) + \ln \left(
  \frac{1-\rho-\eta+\rho'}{\rho-\rho'} \right).
\end{equation}
We then solve $\partial F/\partial\rho=0$ for $\rho'$, which yields
\begin{equation}
\rho'=\frac{\rho\,(c+1)+\eta-1}{1+c},\ \ \text{where}\ \ 
c=\frac{(1-\rho)^2\,(\rho-\beta_{M-1}/2)}{\rho^2\,(1-\rho-\beta_{M-1}/2)}.
\label{eq:delta}
\end{equation}
We can now search for a maximum of $F(\boldsymbol{\gamma},\rho',\eta)$ by using
(\ref{eq:gamma_1}) and (\ref{eq:gamma_l}), for $1\le \ell 
\le M-1$, and (\ref{eq:delta}).

\begin{figure}[htb]
  \centerline{\includegraphics[scale=0.5]{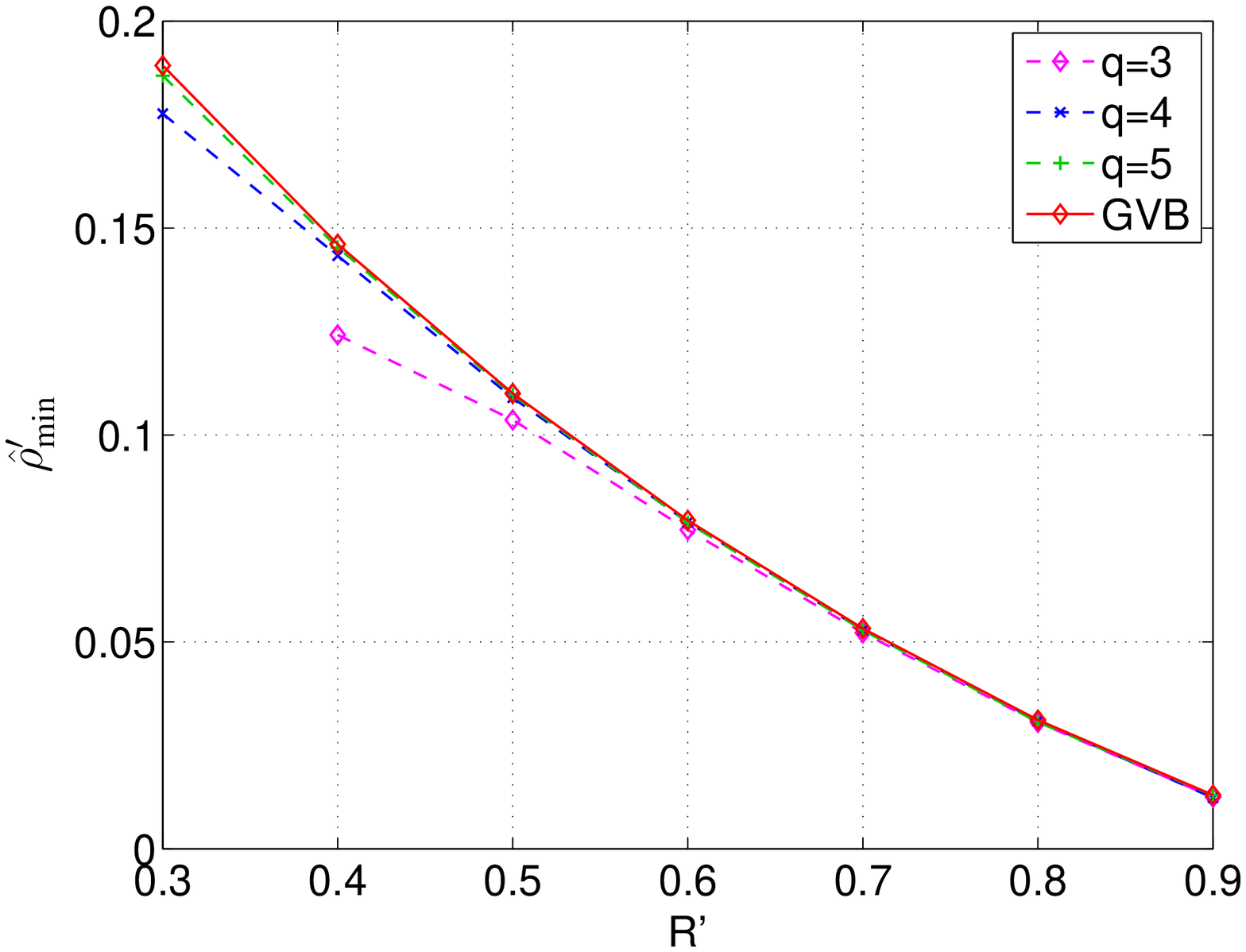}}
  \caption{GVB and the corresponding normalized asymptotic minimum
    distance lower bound $\hat{\rho}_{\text{min}}'$ for the randomly punctured RAA
    code ensemble with mother code  rates $R=1/q=1/3,1/4,1/5$.}
    \label{fig:RAApunct}
\end{figure}

Fig.~\ref{fig:RAApunct} considers the particularly interesting RAA
case and shows the lower bound on the asymptotic minimum distance
growth rate coefficient $\hat{\rho}_{\text{min}}$ for different mother
code rates $R$ and punctured code rates $R'$. We observe that,
compared to the unpunctured RAA code ensemble considered in
Section~\ref{sec:RAA}, the asymptotic minimum distance growth rate
coefficients are closer to the GVB for the punctured ensembles with
rates $R'>R$.  This behavior is due to the extra randomness added by
puncturing the encoder output.  We also see that the growth rate
coefficients approach the GVB as the rate increases.  We conjecture
that this is due to the fact that a smaller value of $\eta$ leads to a
larger random puncturing ensemble. In other words, a smaller $N'$
results in a more "random-like" construction.
Table~\ref{tab:bound_RAA_punct} gives the lower bound on the
asymptotic minimum distance growth rate coefficient
$\hat{\rho}_{\text{min}}'$, along with the estimated values from a
finite-length analysis $\tilde{\rho}_{\text{min}}'$ and the GVB, for a
rate $R=1/3$ mother code.

\begin{table}[htb]
  \caption{Asymptotic minimum distance growth rate coefficient 
  lower bound $\hat{\rho}_{\text{min}}'$, estimated
  minimum distance $\tilde{\rho}_{\text{min}}'$ from a
  finite length analysis,  and the corresponding
  values of the GVB for the punctured RAA
    ensemble with different code rates employing a  mother code of
    rate $1/3$.}
\label{tab:bound_RAA_punct}
\begin{center}
\begin{tabular}{|c|c|c|c|}
\hline
$R$\st  & $\hat{\rho}_{\text{min}}'$ &  $\tilde{\rho}_{\text{min}}'$ & GVB \\\hline \hline
0.4 \st & 0.1242 & 0.1256  & 0.1461   \\\hline
0.5 \st & 0.1036 & 0.1039  & 0.1100 \\\hline
0.6 \st & 0.0771 & 0.0782  & 0.0794 \\\hline
0.7 \st & 0.0522 & 0.0530 & 0.0532 \\\hline
0.8 \st & 0.0306 & 0.0314 & 0.0311 \\ \hline
0.9 \st & 0.0125 & 0.0133 & 0.0130 \\  
\hline
\end{tabular}
\end{center}

\end{table}

The following theorem  is analogous to Theorems~\ref{th:RAA}
and~\ref{th:MRAA}.
\begin{theorem}
\label{th:MRAApunct}
  Consider the ensemble of RMA codes with $M$ accumulators, $M\ge 2$, 
  of rate $R'$ after random puncturing. If the block length $N'\rightarrow\infty$,
  almost all codes in the ensemble have minimum distance $d_{\text{min}}$ 
  growing linearly with $N'$. A lower bound on the asymptotic minimum 
  distance growth rate coefficient $\rho'_{\text{min}}$ of the 
  ensemble can be obtained by solving the system of equations
  \eqref{eq:gamma_1}, \eqref{eq:gamma_l}, and \eqref{eq:delta}, i.e., by
  finding the maximum of the function $F(\boldsymbol{\gamma},\rho',\eta)$. 
\end{theorem}

\section{Concluding remarks}

We have shown that RAA code ensembles for code rates equal to $1/3$ or
smaller are asymptotically good in the sense that their average
minimum distance grows linearly with block length $N$ as
$N\rightarrow\infty$.  Moreover, we have shown that the asymptotic
growth rate coefficients approach the GVB for small code rates.  This
extends the results of \cite{PS03}, where linear distance growth is
only shown for an infinite number of accumulators. These new results
also extend those in \cite{Pfi03} and \cite{BMS03}, where linear
distance growth for the RAA ensemble is shown, but a growth rate
coefficient is either not given or only conjectured.  Similar results
are also obtained for RMA code ensembles with $M \geq 3$ and code
rates equal to $1/2$ or smaller. Further, by introducing random
puncturing at the output of the inner accumulator, we demonstrate that
the resulting high rate RMA ensembles exhibit linear distance growth,
where the asymptotic growth rate coefficient is close to the GVB if
the mother code rate is sufficiently low.

Despite the fact that the RMA code ensembles considered in this paper are
asymptotically good, the convergence behavior of these codes may not 
be sufficient to provide an iterative decoding threshold close to capacity, 
as can be seen from the simulation results presented in \cite{PS03}. 
However, the results obtained may be useful in constructing similar code 
ensembles based on simple component codes with low encoding complexity, 
asymptotically linear distance growth, and good convergence behavior. 
In particular, for the interesting class of double serially concatenated 
codes, the RAA ensemble can be used as a starting point to design asymptotically
good code constructions by replacing one or more of the constituent
encoders with small memory convolutional codes, whose code polynomials
can be chosen to improve iterative decoding convergence behavior. Some
initial experimental results in this direction for different component encoders
are presented in \cite{VKKCZ08,KKZC08}. Another approach is to consider
hybrid concatenated coding schemes, where parts of the encoder are
structurally equivalent to RAA encoders. Initial results for low code
rates have shown that these schemes have improved threshold behavior
compared to RAA codes, while still providing asymptotically linear distance
growth, albeit with a smaller growth rate coefficient \cite{KAKVC08}.

\begin{appendix}

\subsection{Proof of Lemma 1}
\label{sec:proof_lemma1}
We consider the following five partial cases:
\begin{align}
	1-\frac{2}{q} \le a = b \le c < &1 \label{eq:case1} \\
	1-\frac{2}{q} \le a = b < c \le &1 \label{eq:case2} \\
							0 \le a < b \le c < &1 \label{eq:case3} \\
							0 \le a < b  <  c = &1 \label{eq:case4} \\
							0 \le a < b  =  c = &1 \label{eq:case5}.						
\end{align}
From Theorem \ref{th:RA_lowerbound} it  follows that $d_1 \ge \beta\,N^{1-2/q}$ for almost all codes in the ensemble and thus we do not need to consider the case
of $b<1-\frac{2}{q}$.

In addition to \eqref{eq:bcoeff_bound}, we will use the inequality \cite{Gal63}
\begin{equation}
	\sqrt{\frac{N}{8l(N-l)}} \exp \left( H\left( \frac{l}{N}\right)\,N\right)
	\leq {N \choose l} \leq
	\sqrt{\frac{N}{2\pi l(N-l)}} \exp \left( H\left( \frac{l}{N}\right)\,N\right),
	\label{eq:bcoeff_bound2}
\end{equation}
or the equivalent expression
\begin{equation}
	{N \choose l} = \exp \left( H\left( \frac{l}{N}\right)\,N + o(N)\right)
	\label{eq:bcoeff_bound3}
\end{equation}
to bound binomial coefficients.

\begin{itemize}
	\item[(a)] Cases \eqref{eq:case1} and \eqref{eq:case2}: \\
	Using \eqref{eq:bcoeff_bound}, \eqref{eq:cond}, and \eqref{eq:bcoeff_bound2}, we can rewrite 
	$E(A_{d_1,w})={K \choose w}\, \Pr(d_1|w)$ as
	\begin{equation}
		E(A_{d_1,w}) = \left( \frac{N}{q\alpha N^a} \right)^{\alpha N^a} 
						\exp\left( H\left(
                                                    \frac{q\alpha}{2\beta}\right)\,\beta N^a \right) \,
						\left( \frac{2N}{q\alpha N^a} \right)^{\frac{q\alpha}{2} N^a}
						\left( \frac{N}{q\alpha N^a} \right)^{-q\alpha N^a} 
						c_N(w),			
	\label{eq:cwef_a}
	\end{equation}
	where
	\begin{equation}	
		c_N(w) = \frac{\varphi_w(w) \left( \varphi_{\frac{qw}{2}}(\frac{qw}{2})\right)^2 }  							{\varphi_N(qw)}=\exp[o(N^a\,\ln N)]
		\label{eq:c_n}
	\end{equation}
	is a second order term. For simplicity we assume
        $\left\lceil\frac{x}{2}\right\rceil=\left\lfloor\frac{x}{2}\right\rfloor=\frac{x}{2}$,
        which is valid for any even integer $x$ and approximately
        valid for large odd $x$.  Then we have from \eqref{eq:cwef_a}
        that
	\begin{equation}
		E(A_{d_1,w}) = \exp \left[ \left( 1-\frac{q}{2}\right)\alpha \, N^a \ln N
						+ o(N^a\ln N)\right].
	\label{eq:cwef_a2}
	\end{equation}
	Since $E(A_{d_1,w}) \rightarrow 0$ as $N \rightarrow
        \infty$, it follows that $\lim_{N\rightarrow \infty}N^3
        E(A_{d,d_1,w})=0$ independently of $d$ for almost all codes in
        the ensemble.
	
	\item[(b)] Case \eqref{eq:case3}: \\
	To make the expressions more compact, we will omit second order terms from now on.
	Again, using \eqref{eq:bcoeff_bound}, \eqref{eq:cond},  and \eqref{eq:bcoeff_bound2}, we can write
	\begin{equation}
	\begin{aligned}
		E(A_{d_1,w}) \approx & \left( \frac{N}{q\alpha N^a} \right)^{\alpha N^a} 
						 \left( \frac{2\beta N^b}{q\alpha N^a} \right)^{\frac{q\alpha}{2} N^a} \,
						\left( \frac{2N}{q\alpha N^a} \right)^{\frac{q\alpha}{2} N^a}
						\left( \frac{N}{q\alpha N^a} \right)^{-q\alpha N^a}	\\
					=& \exp\left[   \left( (1-a) \left(1-\frac{q}{2}\right)\alpha 
					   + (b-a)\frac{q\alpha}{2} \right) N^a\ln N + o(N^a\ln N)  \right]
	\label{eq:cwef_b}
	\end{aligned}
	\end{equation}
	and
	\begin{equation}
	\begin{aligned}
		\Pr(d|d_1) \approx & \left( \frac{2\rho N^c}{\beta N^b} \right)^{\frac{\beta}{2}N^b} \,
						 						\left( \frac{2 N}{\beta N^b} \right)^{\frac{\beta}{2}N^b} \,
						 						\left( \frac{N}{\beta N^b} \right)^{-\beta N^b} \\
						 				= & \exp \left( -(1-b)\frac{\beta}{2}N^b \ln N 
						 				     + o(N^b \ln N)    \right).
	\end{aligned}					   
	\label{eq:pr_db}
	\end{equation}
	Since $b>a$, it follows from \eqref{eq:cwef_b} and \eqref{eq:pr_db} that
	\begin{equation}
		E(A_{d,d_1,w}) = \exp \left( -(1-b)\frac{\beta}{2}N^b \ln N 
						 				     + o(N^b \ln N)    \right)
	\label{eq:cwef_b2}
	\end{equation}
	and $N^3 E(A_{d,d_1,w}) \rightarrow 0$ as $N \rightarrow \infty$.
	
	\item[(c)] Case \eqref{eq:case4}: \\
	In this case $E(A_{d_1,w})$ is still given by \eqref{eq:cwef_b}, but
	\begin{equation}
	\begin{aligned}
		\Pr(d|d_1) \approx & \left( \frac{2\rho N}{\beta N^b} \right)^{\frac{\beta}{2}N^b} \,
						 					\left( \frac{2 (1-\rho)N}{\beta N^b} \right)^{\frac{\beta}{2}N^b} \,
						 					\left( \frac{N}{\beta N^b} \right)^{-\beta N^b} \\
						 				= & \exp \left( \frac{\beta}{2}N^b \ln (4\rho(1-\rho)) 
						 				     + o(N^b)    \right).
	\end{aligned}					   
	\label{eq:pr_dc}
	\end{equation}
	Since $b>a$, it follows from \eqref{eq:cwef_b} and \eqref{eq:pr_dc} that
	\begin{equation}
		E(A_{d,d_1,w}) = \exp \left( \frac{\beta}{2}N^b \ln (4\rho(1-\rho)) 
						 				     + o(N^b)    \right)
	\label{eq:cwef_c2}
	\end{equation}
	and $N^3 E(A_{d,d_1,w}) \rightarrow 0$ as $N \rightarrow \infty$ for $\rho<1/2$.
	
	\item[(d)] Case \eqref{eq:case5}: \\
	In this case $E(A_{d_1,w})$ is still given by \eqref{eq:cwef_b}, but
	\begin{equation}
	\begin{aligned}
          \Pr(d|d_1) =& \frac{\displaystyle {d \choose \left\lceil
                \frac{d_1}{2}\right\rceil} {N-d \choose \left\lfloor
                \frac{d_1}{2} \right\rfloor} } {\displaystyle {N \choose d_1} } \,
          \frac{ \left\lceil  \frac{d_1}{2}\right\rceil } {d} = \frac{
            \displaystyle {d_1 \choose \left\lceil \frac{d_1}{2}\right\rceil} {N-d_1
              \choose \left\lfloor d-\frac{d_1}{2} \right\rfloor} } {
            \displaystyle {N \choose d} } \,
          \frac{ \left\lceil  \frac{d_1}{2}\right\rceil } {d} \\
          =& \exp \biggr[\biggr( \underbrace{ \beta \ln 2 + H \left(
                \frac{\rho-\beta /2}{1-\beta} \right)(1-\beta) - H \left(\rho
              \right)}_{\triangleq\,F_\rho(\beta)}\,\biggr)\, N + o(N) \biggr].
	\end{aligned}					   
	\label{eq:pr_dd}
	\end{equation}
	The function
	\begin{equation}
		F_\rho(\beta) = \beta \ln 2 + H \left(  \frac{\rho-\beta/2}{1-\beta}  \right)(1-\beta)
							 						- H \left(\rho \right)
	\end{equation}
	is strictly negative for all $\rho<1/2$ and $\beta \leq 2\rho$, which can be
	shown as follows.	The derivative of $F_\rho(\beta)$,
		\begin{equation}
		\begin{aligned}
			\frac{\partial F}{\partial \beta} = &
					\frac{1}{2} \ln \left(  \frac{\rho-\beta/2}{1-\beta}  \right)
					+ \frac{1}{2} \ln \left(  \frac{1-\rho-\beta/2}{1-\beta}  \right)
					+ \ln 2 \\
			 =& \frac{1}{2} \ln \left( 4x(1-x) \right),
		\end{aligned}
		\end{equation}
		is negative if $x=\frac{\rho-\beta/2}{1-\beta} \neq 1/2$ and equals zero if
		$x=1/2$. $F_\rho(\beta)$ is thus negative for all $\rho \le 1/2$ and all
		$0 < \beta \leq 2\rho$.

	From \eqref{eq:cwef_b} and \eqref{eq:pr_dd} follows that
	\begin{equation}
		E(A_{d,d_1,w}) = \exp \left[ \left( \beta \ln 2 
							 						+ H \left(  \frac{\rho-\beta/2}{1-\beta}  \right)(1-\beta)
							 						- H \left(\rho \right) \right)N + o(N) \right]
	\label{eq:cwef_d2}
	\end{equation}
	and $N^3 E(A_{d,d_1,w}) \rightarrow 0$ as $N \rightarrow \infty$ for all $\rho<1/2$.
\end{itemize}

\subsection{Proof of Lemma 4}
\label{sec:proof_lemma4}
\begin{itemize}
\item[(a)] On the boundary $(\alpha=0,0\le\beta \le 2\rho)$, we have 
\begin{align}
\frac{\partial f}{\partial \beta}\biggr|_{\alpha=0}&=\frac{1}{2} \ln \left(
  \frac{\rho-\beta/2}{1-\beta}  \right) + \frac{1}{2} \ln \left(
  \frac{1-\rho-\beta/2}{1-\beta}  \right) + \ln 2\notag\\
&= \frac{1}{2} \ln\left( 4\,x_\beta\,(1-x_\beta) \right)<0 \label{eq:dfdbeta1}
\end{align}
where $x_\beta =\frac {\rho-\beta/2}{1-\beta}$. Since
$f(\alpha,\beta,\rho)\big|_{\alpha=0,\beta=0} =0$,  we obtain
$f(\alpha,\beta,\rho)<0$ on the boundary (a) for all $\beta>0$ and
$\rho \leq 1/2$.

\item[(b)] On the  boundary  $(0 \leq \alpha \le
\min(1,4\rho), \beta =\alpha /2)$ the total derivative with respect to
$\alpha$ is given as 
\begin{align}
  \frac{df}{d\alpha}\biggr|_{\beta=\alpha/2}=& \frac{\partial
    f}{\partial \alpha}\biggr|_{\beta=\alpha/2}+\frac{\partial
    f}{\partial
    \beta}\biggr|_{\beta=\alpha/2}\,\frac{d\beta}{d\alpha}\notag \\
 =& -\frac{1}{q}\,\ln\left( \frac{\alpha}{1-\alpha}\right) + \ln 2
 +\frac{1}{2} \, \ln\left( \frac{\alpha/2}{1-\alpha/2} \right)  \notag \\
  &+ \frac{1}{2} \left[ \frac{1}{2}
 \, \ln\left( \frac{\rho-\beta/2}{1-\beta} \right) +\frac{1}{2}
 \, \ln\left( \frac{1-\rho-\beta/2}{1-\beta} \right) +\ln 2 \right]. 
 \label{eq:dfdalpha}
  \end{align}
Note that the inequality
\begin{equation}
\frac{1}{q} \, \ln\left( \frac{\alpha}{1-\alpha} \right) > \frac{1}{2}
 \, \ln\left( \frac{\alpha/2}{1-\alpha/2} \right) + \ln 2
\label{eq:ineq1}
\end{equation}
holds for $q\ge 3$ and $0\le \alpha < 1$. The last term on the
right hand side of \eqref{eq:dfdalpha} is equivalent to  \eqref{eq:dfdbeta1}.
By inserting \eqref{eq:dfdbeta1} and  \eqref{eq:ineq1} into \eqref{eq:dfdalpha}
we obtain $df/d\alpha \bigr|_{\beta=\alpha/2}<0$. Therefore, we 
conclude that $f(\alpha,\beta,\rho)$ is negative along the boundary 
$(0\le \alpha\le \min(1,4\rho),\beta=\alpha/2)$ for all $0\le \rho <0.5$.
\end{itemize}

\end{appendix}

\bibliographystyle{IEEEbib}
\bibliography{lit}

\end{document}